\documentclass[twocolumn,showpacs,preprintnumbers,amsmath,amssymb]{revtex4}

\usepackage{graphicx}
\usepackage{dcolumn}
\usepackage{bm}
\usepackage{soul}
\usepackage{color}
\usepackage{epstopdf}

\begin{document}

\title{van der Waals interactions between graphitic nanowiggles}

\author{ Anh D. Phan$^{1,2}$, Lilia M. Woods$^{1}$, The-Long Phan$^{3}$}
\affiliation{$^{1}$Department of Physics, University of South Florida, Tampa, Florida 33620, USA}%
\email{anhphan@mail.usf.edu}
\affiliation{$^{2}$Institute of Physics, Vietnam Academy of Science and Technology, 10 Dao Tan, Hanoi 10000, Vietnam}%
\affiliation{$^{3}$Department of Physics, Chungbuk National University, Cheongju 361-763, Korea}
\email{ptlong2512@yahoo.com}

\date{\today}

\begin{abstract}
The van der Waals interactions between two parallel graphitic nanowiggles (GNWs) are calculated using the coupled dipole method (CDM). The CDM is an efficient and accurate approach to determine such interactions explicitly by taking into account the discrete atomic structure and many-body effect. Our findings show that the van der Waals forces vary from attraction to repulsion as nanoribbons move along their lengths with respect to each other. This feature leads to a number of stable and unstable positions of the system during the movement process. These positions can be tuned by changing the length of GNW. Moreover, the influence of the thermal effect on the van der Waals interactions is also extensively investigated. This work would give good direction for both future theoretical and experimental studies.
\end{abstract}

\maketitle

\section{Introduction}

The van der Waals interaction, arising from the quantum fluctuations within bodies, has received enthusiastic attention in recent years due to its important role from condensed matter physics \cite{1,2,3} to biological systems \citep{4,5}. At nanoscale, this force is much more dominant than other forces such as the electromagnetic force \citep{6}. In the field of material science, the van der Waals force mainly causes the stiction and friction problems which directly impact behaviors and physical properties of nanodevices. The understanding of such interactions has provided the cornerstone towards next-generation devices and advancement fundamental science.

There are several main ways to estimate the van der Waals interactions. The well-known method implemented in molecular dynamics (MD) simulation codes is based on the Lennard-Jones (LJ) model \cite{7,8,9}. This calculation, however, considers only the pairwise interaction between two atoms and ignores the effects induced by other atoms. Another approach is built up from the Lifshitz theory \citep{10,11,12}. This theory allows us to calculate the van der Waals and Casimir force between two parallel semi-infinite plates using dielectric functions of objects as a function of the imaginary frequency. The interaction between two objects with arbitrary geometries such as sphere-sphere and cylinder-cylinder systems can be calculated by the proximity force approximation (PFA) \cite{10,13}. A different heuristic approach for studying dispersion interactions derives from the multipole expansion \cite{14}. In the framework of this method, the van der Waals energy between two spherical bodies with the centre-to-centre distance $R$ are expressed by
\begin{eqnarray}
E_{vdW} = -\frac{C_6}{R^6}-\frac{C_8}{R^8}-\frac{C_{10}}{R^{10}}-....,
\label{eq:1}
\end{eqnarray}
where $C_6$, $C_8$, and $C_{10}$,... are the van der Waals coefficients that can be obtained by using the dynamical polarizability of each object. To apply Eq.(\ref{eq:1}) effectively, the separation of two spheres is required to be larger than a certain minimum value. In practical complex systems such protein, DNA and graphene nanoribbons, the anisotropy and surface roughness must be taken into account in the van der Waals interactions. As a result, the long-range interaction at short distances estimated by these two above methods may not accurate. In this paper, we use the CDM to calculate the van der Waals interaction energy. This method is based on the many-body effects and has been used broadly in a wide range of research areas \cite{15,16,17}. The CDM is capable of dealing with systems containing a large number of atoms and the effects caused by the structure orientations.
\begin{figure}[htp]
\includegraphics[width=8.5cm]{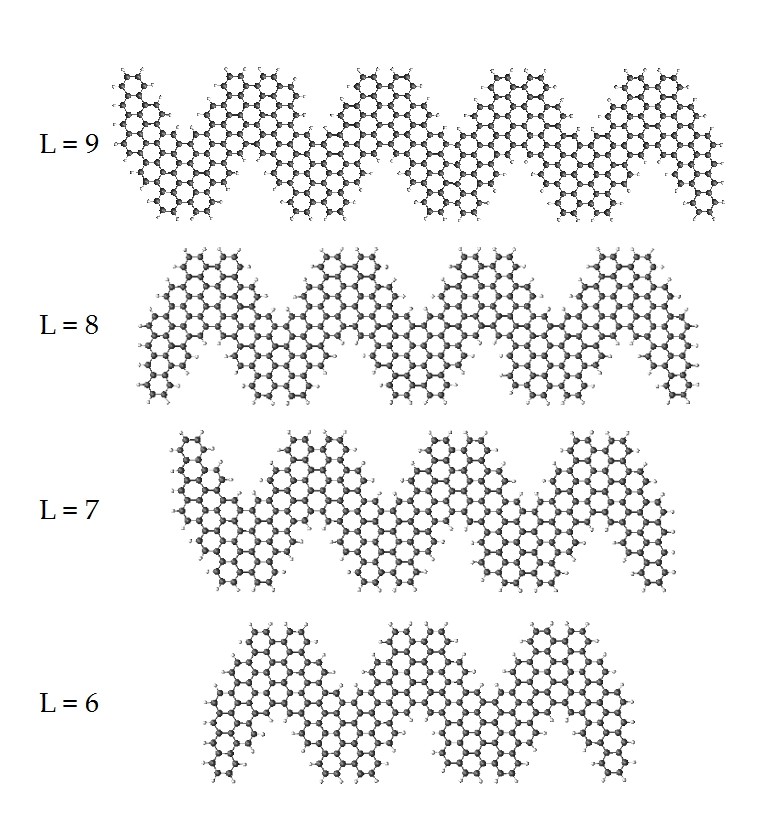}
\caption{\label{fig:1}(Color online) The atomic arrangements of a graphene nanowiggle. The grey circles denote carbon atoms while the white circles refer to hydrogen atoms. The length of GNW is specified by the number of $L$.}
\end{figure}

It is well-known that graphene is an extraordinary material due to its fascinating properties and possible applications \cite{18}. Graphene also provides state-of-the-art classes of material such as carbon nanotubes (CNTs), bulky balls, graphene nanoribbons (GNRs) and GNWs (see Fig.\ref{fig:1}). Recently, both experimentalists \cite{19,20} and theoreticians \cite{21,22,23} have been attracted by GNWs. The finite-size effects and the modification of the edge in the chevron nanostructure influence significantly its properties and lead to be better performance than pristine graphene. The increase of electron-hole confinements strengthens the exciton binding energy over that of regular GNRs \cite{23}. One can take advantage of this feature by designing optical sensors to be used in biosystems \cite{24}. Moreover, it was shown that the chevron-like GNRs have large values of the figure of merit and can be useful in thermoelectric devices \cite{22}. In this work, we focus on the van der Waals interaction in chevron-like GNRs.

\section{Theoretical background}
This computational theory starts by considering a system as set of N discrete polarizable atoms interacting to each other via electromagnetic fluctuations. A polarization of the dipole $\mathbf{p}_i$ at position $\mathbf{r}_i$ is induced by \cite{26}
\begin{eqnarray}
\mathbf{p}_i=\mu_0\alpha_i\omega^2\sum_{j\neq i}^N\left(\mathbb{G}_{ij}\mathbf{p}_j + i\frac{\mathbf{g}_{ij}\times \mathbf{m}_j}{\omega}  \right),
\label{eq:2}
\end{eqnarray}
where $\mu_0$ is the magnetic permeability of the vacuum, $\mathbf{m}_j$ is the magnetic dipole moment of the \emph{j}th dipole, the free-space Green's function is given 
\begin{eqnarray}
\mathbb{G}_{ij}=\frac{e^{ikR_{ij}}}{4\pi R_{ij}}\left[\left(1+\frac{i}{kR_{ij}}-\frac{1}{k^2R_{ij}^2}\right)\mathbb{I} \right.\nonumber\\
\left. + \left(-1-\frac{3i}{kR_{ij}} + \frac{3}{k^2R_{ij}^2}\right) \right]\hat{R}_{ij}\otimes \hat{R}_{ij},
\label{eq:12}
\end{eqnarray}
where $k = \omega /c$ is is the wave number, c is the speed of light, $\mathbb{I}$ is the $3 \times 3$ identity matrix, $\mathbf{R}_{ij}=\mathbf{r}_{i}-\mathbf{r}_{j}$ is the distance vector between dipole i and j, $\hat{R}_{ij}=\mathbf{R}_{ij}/R_{ij}$, and the vector $\mathbf{g}_{ij}$ is expressed
\begin{eqnarray}
\textbf{g}_{ij}=\frac{e^{ikR_{ij}}}{4\pi}\left(\frac{ik}{R_{ij}}-\frac{1}{R_{ij}^2}\right)\hat{R}_{ij}.
\label{eq:13}
\end{eqnarray}

For short separations $kR_{ij} \ll 1$, the retardation effects can be ignored. In this range, the highest order terms of $1/R$ dominate over others. The magnetic contributions in Eq.(\ref{eq:2}), therefore, is neglectable. Eq.(\ref{eq:2}) can be rewritten as $\mathbf{p}_i = \mu_0\alpha_i\omega^2\sum_{j\neq i}\mathbb{T}_{ij}\mathbf{p}_j$. Here
\begin{eqnarray}
\mathbb{T}_{ij} = \frac{3}{4\pi k^2R_{ij}^5}
\left[ \begin{array}{ccc}
x_{ij}^2 & x_{ij}y_{ij} & x_{ij}z_{ij}  \\
y_{ij}x_{ij} & y_{ij}^2 & y_{ij}z_{ij}  \\
z_{ij}x_{ij} & z_{ij}y_{ij} & z_{ij}^2
\end{array} \right] -\frac{1}{4\pi k^2 R_{ij}^3}\mathbb{I},
\label{eq:3}
\end{eqnarray}
in which $\mathbf{x}_{ij}$, $\mathbf{y}_{ij}$ and $\mathbf{z}_{ij}$ are components along $x$, $y$ and $z$ axis of $R_{ij}$, respectively. It is important to note that $\mathbb{T}_{ij} = 0$ as $i = j$. $\alpha_i$ is the atomic polarizability of dipole $i$ modelled by the Drude model \cite{25}
\begin{eqnarray}
\alpha_i(\omega)=4\pi\varepsilon_0\frac{\alpha_{0i}\omega_{0i}^2}{\omega_{0i}^2-\omega^2},
\label{eq:4}
\end{eqnarray}
where $\alpha_{0i}$ is the static polarizability and $\omega_{0i}$ is the characteristic frequency of the \emph{i}th dipole. In our work, the atomic polarizability is assumed to be isotropic. Now it is easy to obtain the 3N coupled equations \cite{29}
\begin{eqnarray}
\left[ \begin{array}{cccc}
\mathbb{I}/\alpha_{01} & -\mathbb{T}_{12} & \ldots & -\mathbb{T}_{1N} \\
-\mathbb{T}_{21} & \mathbb{I}/\alpha_{02} &\ldots & -\mathbb{T}_{2N}  \\
\vdots & \vdots & \vdots & \vdots \\
-\mathbb{T}_{N1} & \ldots & \ldots & \mathbb{I}/\alpha_{0N}
\end{array} \right]
\left[ \begin{array}{c}
\mathbf{p}_1 \\
\mathbf{p}_2\\
\vdots \\
\mathbf{p}_N
\end{array} \right] = \omega^2 \times    \nonumber\\
\left[ \begin{array}{cccc}
\mathbb{I}/(\alpha_{01}\omega_{01}^2) & 0 & 0 \\
0 & \mathbb{I}/(\alpha_{02}\omega_{02}^2) &\ldots & 0  \\
\vdots & \vdots & \vdots & \vdots \\
0 & \ldots & \ldots & \mathbb{I}/(\alpha_{0N}\omega_{0N}^2) 
\end{array} \right]
\left[ \begin{array}{cccc}
\mathbf{p}_1 \\
\mathbf{p}_2\\
\vdots \\
\mathbf{p}_N
\end{array} \right]. 
\label{eq:6}
\end{eqnarray}

Solving Eq.\eqref{eq:6} provides $3N$ values of $\omega$. The van der Waals interactions of the entire system at 0 $K$ is expressed by
\begin{eqnarray}
U^{vdW}=\sum_{i=1}^{3N}\frac{\hbar\omega_i}{2}.
\label{eq:7}
\end{eqnarray}

The retarded dispersion energy between object 1 and 2 are calculated by
\begin{eqnarray}
U^{vdW}_T=U^{vdW}- U^{vdW}_1-U^{vdW}_2,
\label{eq:8}
\end{eqnarray}
where $U^{vdW}_1$, $U^{vdW}_2$ and $U^{vdW}$ are the van der Waals energy of only object 1, object 2 and the whole system including both of them, respectively. In some cases, the tensor $\mathbb{T}_{ij}$ given by Eq.\eqref{eq:3} leads to a ``polarization catastrophe" problem. To avoid this issue, it is necessary to introduce the distance-dependent function $s_{ij}=a\left(\alpha_{0i}\alpha_{0j}\right)^{1/6}$ and modify the coupled dipole interaction tensor \cite{30,31}
\begin{eqnarray}
\mathbb{T}_{ij} = \frac{3\upsilon_{ij}^4}{4\pi k^2 R_{ij}^3}\hat{R}_{ij}\otimes \hat{R}_{ij}
- \frac{(4\upsilon_{ij}^3-3\upsilon_{ij}^4)}{4\pi k^2 R_{ij}^3}\mathbb{I},
\label{eq:9}
\end{eqnarray}
where $a = 1.662$ is the screening factor. If $R_{ij} > s_{ij}$ then $\upsilon_{ij} = 1$, else if $R_{ij} < s_{ij}$ then $\upsilon_{ij} = R_{ij}/s_{ij}$.
\section{Numerical results and discussion}
In this section, we investigate the van der Waals interactions between two chevron-like GNRs in vacuum. The used parameters for atomic polarizabilities \cite{29,32} are $\omega_C=1.85\times10^{16}$ $rad/s$, $\omega_H=1.41\times10^{16}$ $rad/s$, $\alpha_C=0.85$ $\AA^3$, and $\alpha_H=0.25$ $\AA^3$ . Each sheet is relaxed by molecular dynamics (MD) simulation ReaxFF and then we combine these two ribbons to create a static system.

\begin{figure}[htp]
\includegraphics[width=8cm]{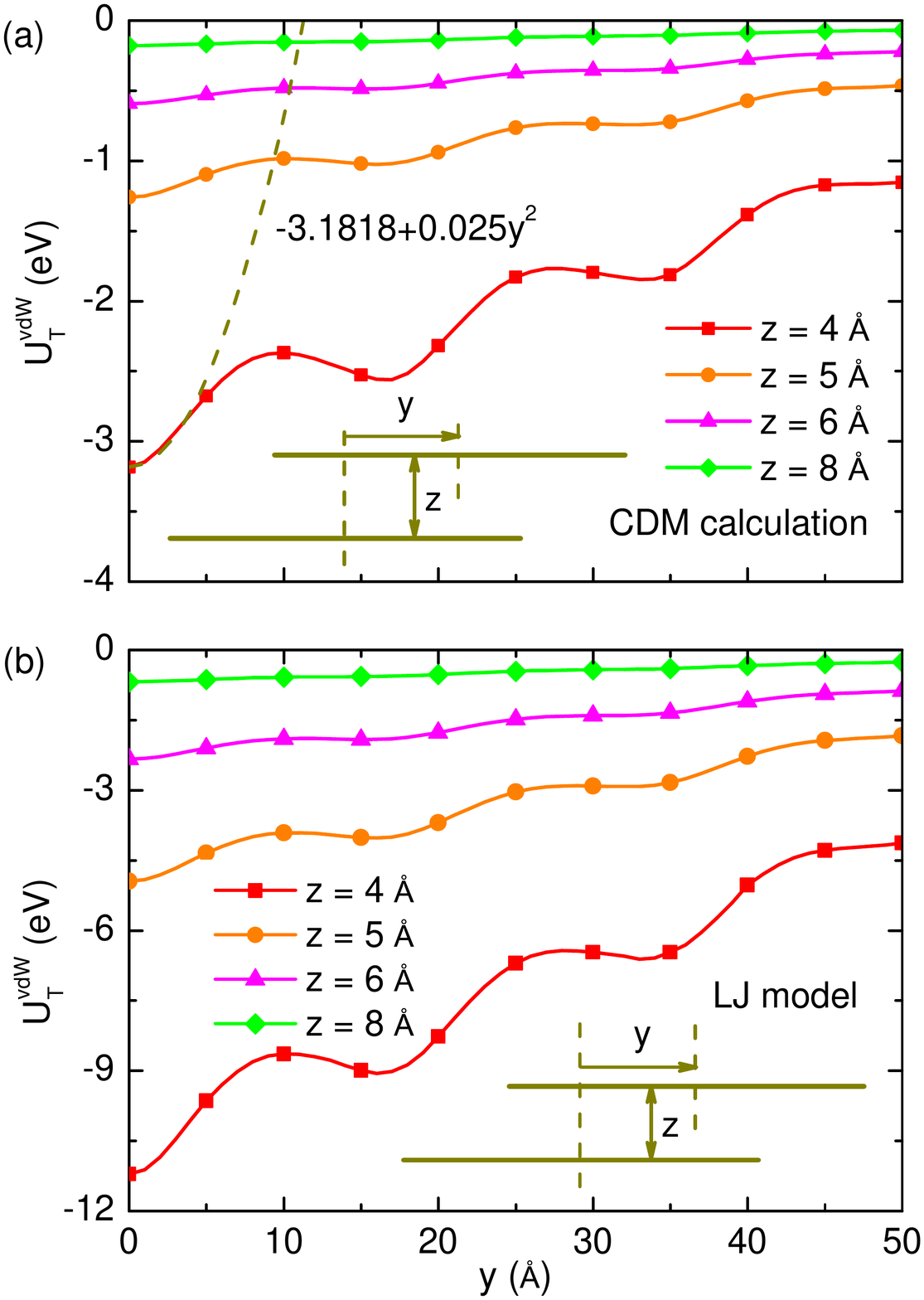}
\caption{\label{fig:2}(Color online) The van der Waals interactions of two GNWs versus the movement of top nanoribbon along the $y$ axis at 0 $K$. The dashed line illustrates the fitting function of the van der Waals energy around y=0 at z = 4 $\AA$.}
\end{figure}

Figure \ref{fig:2}a shows the dispersion energy between two parallel identical chevron-like GNRs with $L = 9$ at several certain separation distances z along the $z$ axis using the CDM method. As the separation z increases at the same y, this interaction decreases significantly. Drifting the top nanoribbon along the $y$ axis causes an oscillation of the retarded dispersion energy. This oscillatory-like feature reveals that the van der Waals force changes from attraction to repulsion and vice versa. At z = $4$ $\AA$, the oscillation is so strong, and we can see the stable equilibrium positions of the system at local minimum energies corresponding to y = $0$, 18 and 34 $\AA$. Obviously, the most stable equilibrium state is at y $= 0$ for all distance z, other positions have feeble barriers and can be called the metastable positions. At long separation distances, the oscillatory-like properties of energy reduces and almost disappears at z = 8 $\AA$. Because the influence of atomic structures, particularly stacking, is significantly weakened when the separation distance z is beyond the C-C bond length.

The results in Fig.\ref{fig:2}a also present that the oscillatory trend between two nearest peaks is approximately 18 $\AA$. This distance corresponds to the salient geometrical period of the chevron-like GNRs depicted in Fig.\ref{fig:1}. The interpretation for this peculiarity is that the $x-y$ plane projection of the top nanoribbon match with that of the bottom one. The oscillatory feature can obtained using the Lennard-Jones (LJ) 12-6 potential $V_{ij}=-A/R_{ij}^6+B/R_{ij}^{12}$ shown in Fig.\ref{fig:2}b. Here A and B are the Hamaker constants. The parameters $A_{CC} = 15.2$ eV$\AA^6$ and $B_{CC}=24\times 10^3$ eV$\AA^{12}$ representing the C-C interaction in the graphene-graphene system was introduced in Ref.\cite{28}. The parameters for the C-H and H-H interactions $A_{CH}=11.4$ eV$\AA^6$, $B_{CH}=11.77\times 10^3$ eV$\AA^{12}$, $A_{HH}=8$ eV$\AA^6$ and $B_{HH}=5.05\times 10^3$ eV$\AA^{12}$ were calculated from the MD simulation such as LAMMPS code. Although the L-J model is the simplest method to estimate the interaction between two neutral atoms and has been widely used in numerous MD codes, the pairwise approach does not provide highly accurate results in comparison with the CDM due to the underestimation of the many-body effects. Our findings and previous study \cite{35} show the difference of the magnitude of the van der Waals interactions derived from the L-J model and the CDM method. One, however, can observe the same phenomenon by these two calculations.   

Using fitting function, the approximate analytical expression of the van der Waals energy $U_{fit}$ around z = 4 $\AA$ and y = 0 is given in Fig. \ref{fig:2}a. The form of this fitting function indicates that we can consider the small movement of top GNR along the $y$ axis as harmonic motion. The frequency of this oscillation can be easily calculated by 
\begin{eqnarray}
\nu_0 = \frac{1}{2\pi}\sqrt{\frac{1}{m}\frac{d^2U_{fit}}{dy^2}},
\label{eq:10}
\end{eqnarray}
where m is the mass of the top nanoribbon. One can estimate $m=3\sqrt{3}na^2\rho_g/2$, in which $a=1.42$ $\AA$ is the bond length between two carbon atoms, $\rho_g=7.6\times10^{-7}$ $kg/m^2$ is the surface mass density of graphene \cite{33}, and $n=132$ is the number of hexagon in GNWs. Substituting all parameters to the expression of $m$, we obtain $m=5.255\times 10^{-24}$ $kg$ and the frequency $\nu_0 \approx 2.42\times 10^{11}$ $Hz$. Another method to calculate $m$ is the sum of mass of individual atoms in the ribbon \cite{34} and this approach leads to $m= 7.76\times 10^{-24}$ $kg$. However, this estimation does not take into account the reduction of total mass due to the formation of the binding energy when bringing isolated atoms closer together. At retarded distances, the van der Waals interactions dominates other forces. The motion of the top flake is controlled only by this long range interactions.

\begin{figure}[htp]
\includegraphics[width=9.5cm]{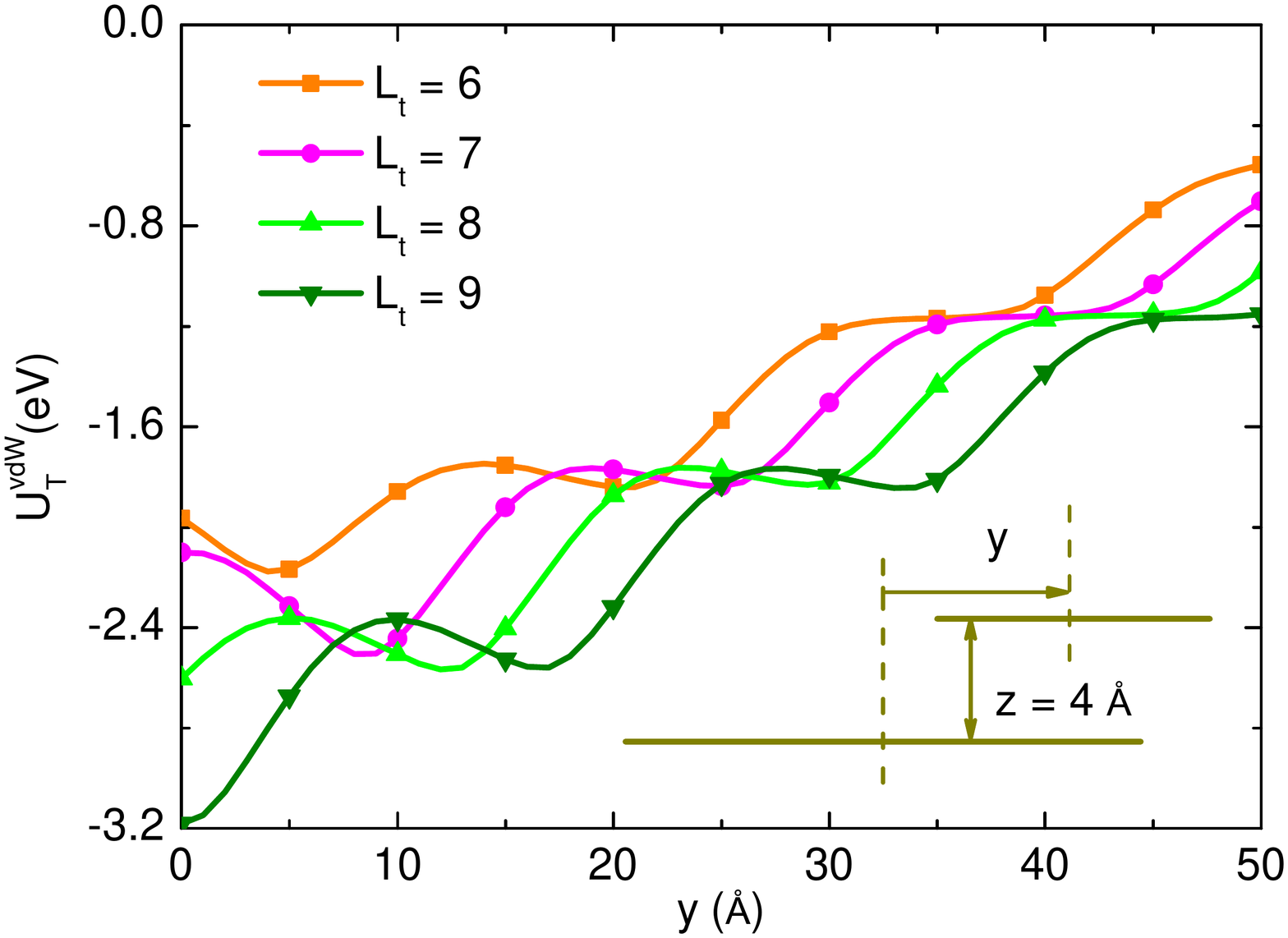}
\caption{\label{fig:3}(Color online) The van der Waals interactions of two GNWs versus the movement of top nanoribbon along the $y$ axis.}
\end{figure}

In the next step, we study the effect of GNW length on the van der Waals energy. The length of bottom GNW remains unchanged $L_{b} = 9$ while the length of top GNW varys with $L_{t}$ = 6, 7, 8 and 9. As can be seen in Fig. \ref{fig:3}, the dispersion interactions between two GNWs at the same separation z = 4 $\AA$ the reduction of $L_{t}$ leads to the unbalance of top nanoribbons at y = 0. The stable equilibrium positions are approximately 8 $\AA$ and 4 $\AA$ at $L_{t} = 7$ and 6, respectively. For $L_{t}=8$, the local minima of the energy interactions at y = 0 and 12.5 $\AA$ have the same energy value, however, the energy barriers around y is smaller than those of y = 12.5 $\AA$. Therefore, y = 12.5 $\AA$ would be the most stable equilibrium position of this system. Our findings indicate that the finite size structure and the edge state have an immense impact on the van der Waals energy. Furthermore, there are a number of repulsive-attractive transitions of the van der Waals force at y $\le 36$ $\AA$. At larger distances, only attractive van der Waals force exists.

It is remarkable that one can modify Eq.(\ref{eq:7}) to estimate the van der Waals energy of the whole at certain temperature $T$
\begin{eqnarray}
U_{tem}^{vdW}= \sum_{i=1}^{3N}\frac{\hbar\omega_i}{2}\coth\left(\frac{\hbar\omega_i}{2k_BT}\right),
\label{eq:11}
\end{eqnarray}
where $k_B$ is the Boltzmann constant. At room temperature, $k_BT \approx 0.025$ $eV$ and all eigenvalues provide $\hbar\omega_i /k_BT \gg 1$. As a result, Eq.(\ref{eq:7}) can be applicable to the calculation at non-zero temperature in this case, so $U_{tem}^{vdw}=U_T^{vdw}$ in terms of formula. The interesting problem is that the thermal effect influences the roughness of the nanowiggle. The different rough surfaces result in different energy interactions.
\begin{figure}[htp]
\includegraphics[width=9.5cm]{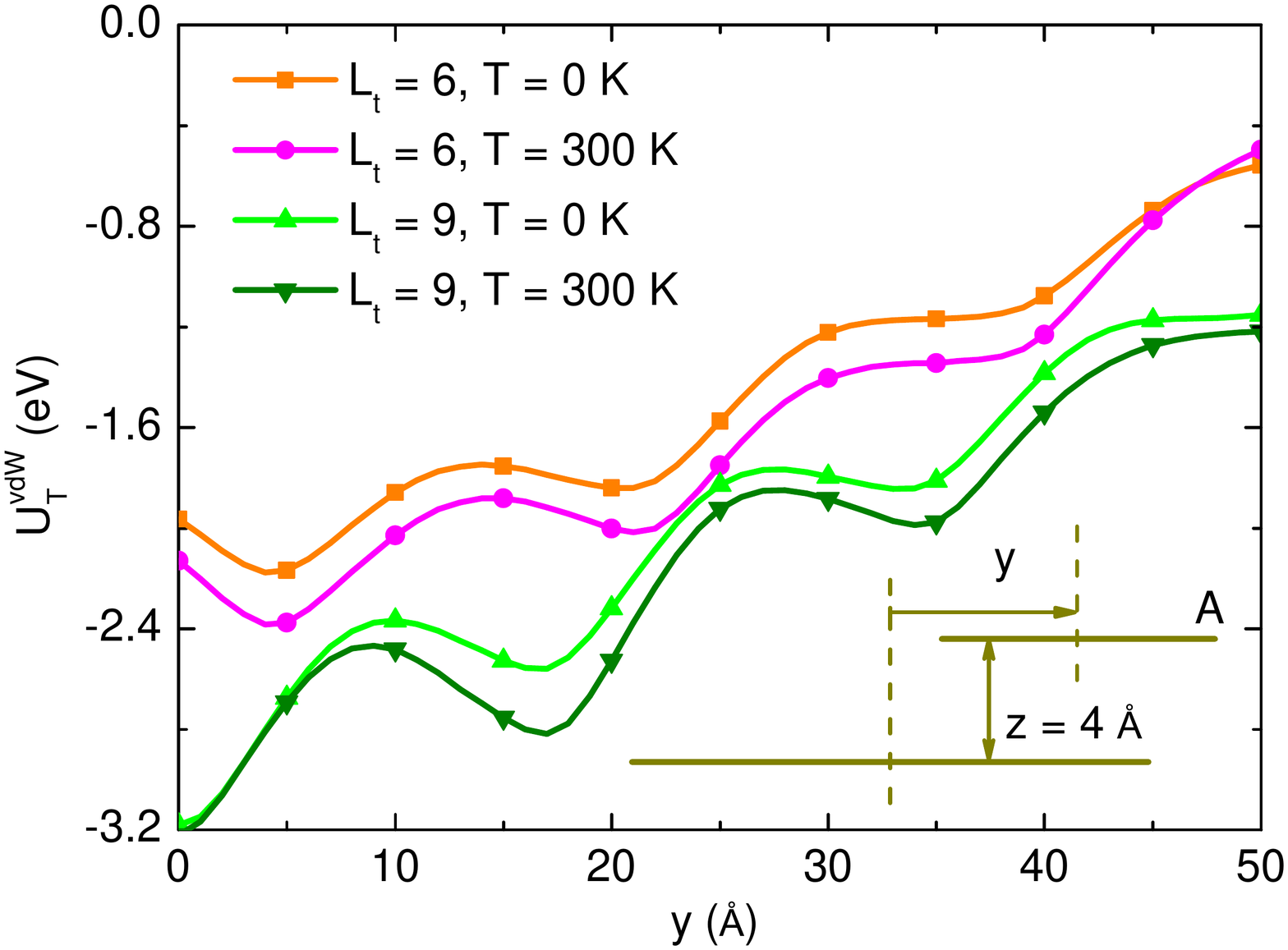}
\caption{\label{fig:4}(Color online) The van der Waals interactions of two GNWs versus the movement of top nanoribbon along the $y$ axis at zero and room temperature when the center-center distance of two objects in $z$ direction is 4 $\AA$.}
\end{figure}

To understand the impact of the thermal effect on the van der Waals energy, we calculate the interactions similar way to previous calculations, however, each GNR is relaxed separately by ReaxFF program at 300 $K$. It can be seen in Fig. \ref{fig:4}, the van der Waals energy interactions in the system including the small GNW are more sensitive than systems possessing the longer chevron-like GNR. It is well-known that the bond in edge is weaker compared to the rest area on the flake. Interestingly, temperature induces the bend of the GNW edge. This deformation generates the energy difference in the system under different thermal conditions. For short nanoribbons, the edge effect plays an important role in the interaction between two objects. This effect, however, has less contribution as the length of GNW is increased. For GNW $L_t = 9$, the dispersion energy is nearly unchanged at y $\le$ 10 $\AA$ as temperatures vary. While one can observe the large discrepancy of such interaction as $T$ = 0 $K$ and $T$ = 300 $K$. At large y, the energy difference at different temperature in both cases $L_t = 6$ and $L_t = 9$ tend to disappear. This result implies that the edge A is far apart enough to ignore the influence of edge atomic coordinates on the van der Waals interactions.

\section{Conclusions}
We have discussed the van der Waals interaction between two chevron-like GNRs by means of CDM. The influence of stacking and edge effect causes the oscillatory-like feature of this energy as the top flake moves parallel to its length direction. Local minima and maxima of energy curve is shifted when the length of a nanoribbon is varied. This coincides with the change of the most stable position of the system. Heating up systems induces the variation of edge atomic positions. At retarded distances, this structure change provides the difference of van der Waals energy at different temperatures. As the center-center separation of two GNWs increases, the energy discrepancy becomes smaller.

\begin{acknowledgments}
Lilia M. Woods acknowledges the Department of Energy under contract DE-FG02-06ER46297.
\end{acknowledgments}

\end{document}